# Meteorological assessment of vertical axis wind turbine energy generation potentials across two Swiss cities in complex terrain

**Aldo Brandi[a,*], Gabriele Manoli[a]**

[a] Laboratory of Urban and Environmental Systems, École Polytechnique Fédérale de Lausanne, Switzerland

*Corresponding author: Aldo Brandi (aldo.brandi@epfl.ch)
BP 3137 (Bâtiment BP), Station 16, CH-1015 Lausanne, Switzerland
ORCID: 0000-0003-2676-7106

## Abstract

Wind energy is the most mature renewable energy technology, however, its exploitation in cities is often met with skepticism. Thanks to their ability to operate effectively at low wind from any direction, vertical axis wind turbines (VAWTs) offer an attractive opportunity for wind energy harvesting in cities, but limited evidence exists on their potential in complex urban environments, and the role of different geographical settings, local meteorological conditions, and urban characteristics remains unclear. Here we use realistic Weather Research and Forecast model high-resolution wind speed simulations alongside representative VAWT power curves to quantify the range of micro-generation potentials at the annual, seasonal, and diurnal scale across two Swiss cities (Lausanne and Geneva) residing in complex terrain. Our results show that Lausanne generally experiences higher (+24%) wind speed than Geneva. Both cities present the greatest micro-generation potential during the summer months, although Lausanne shows non-negligible potential also during the wintertime. Wind speed is higher during the nighttime in Lausanne and during the daytime in Geneva, due to the different interaction between the local lake-breeze circulation and the synoptic flow. Simulated performance of case-study VAWTs is dominated by cut-in wind speed and power curve inflection point. On average, in 2022, an individual VAWT would have produced 2665 kWh of total annual electricity, equivalent to 16.5 square meters of photovoltaic panels. These results highlight the need for research on urban wind energy featuring detailed city-scale assessments that account for urban heterogeneities and regional circulation patters, to inform future planning investment and engineering development.

# 1. Introduction

Urban environments account for the majority of global energy consumption [1] and greenhouse gas (GHG) emission [2]. The environmental, health, and economic impacts of fossil fuel-driven climate change highlight the need to expand and accelerate the transition to renewable energy (RE) globally, and for urban environments to be at the forefront of such a transformation. The challenge is compounded by projections of increasing urban population, and associated energy demand [3]. Therefore, all available options ought to be taken into consideration and their potential systematically explored.

Wind energy is the most mature of RE technologies, and it currently offers the best balance between power generation and environmental impact [4]. In recent years, global wind energy installed capacity has been increasing on average at a 10% annual rate, totaling 976 GW in June 2023 [5]. However, these figures refer to large scale wind farms sited away from urban settlements, as the level of implementation of wind turbines in urban environments remains negligible and no data on wind energy micro-generation in urban areas currently exist. Diversifying the spatial distribution of wind turbines (WTs) across natural and urban environments presents several advantages. The main one is the reduction of energy losses due to transport and distribution. Another considerable advantage is the optimization of residential RE production, currently consisting of photovoltaic (PV) panels only, as wind energy generation can happen throughout the entire diurnal cycle and during the wintertime when photovoltaic output is usually low. In addition, the impact on natural ecosystems, with the associated loss of biodiversity [4] can be reduced by shifting part of the needed deployment on already anthropized landscapes.

Cities have traditionally been considered unsuitable environments for wind energy harvesting, as their distinctive three-dimensional morphology disrupts the wind flow near the surface and decrease wind speed [6,7]. The drag effect exerted by buildings adds to the wind shear component of turbulence within the Urban Boundary Layer (UBL) and its impact can extend up to three times their vertical dimension [8]. On the other hand, enhanced convective mixing typical of urban environments can attenuate the drag effect of built surfaces [9,10] and some studies even suggested the potential for an urban wind island effect, as wind flow can accelerate over the city at specific background meteorological conditions [11,12]. The well-known urban heat island (UHI) effect, defined as the surface or air temperature difference between urban and rural areas driven by the abundance of paved surfaces and anthropogenic heat of cities [13], can generate pressure gradients across landscapes and induce local circulations in the form of country breeze flows [10]. Moreover, cities located in complex terrain are affected by topographical flows, like valley and slope winds, whose impacts on air quality and thermal comfort are still being investigated [14]. This degree of complexity affects the quantification of wind resources in urban areas, and the idea of harvesting wind energy within cities is often met with skepticism.

Thanks to engineering advancements in vertical axis wind turbine (VAWT) technologies, there are currently several commercially available options of small, lightweight VAWTs that can easily be installed within urban fabrics (e.g., on building rooftops or within public spaces) and produce up to several kWh of electricity. VAWTs have several qualities that make them ideal for use in urban environments: they are more efficient in weak wind conditions thanks to lower cut-in wind speeds (i.e., the threshold above which wind turbines start spinning and producing electricity) and better performances in turbulent flow conditions, they operate regardless of wind direction, and result in lower noise levels compared to horizontal axis WTs. Small size VAWTs can be installed along major roadways to harvest the turbulent wind produced by vehicle traffic and, when combined to solar panels, could be used to power distributed off-grid charging stations for electric vehicles.

Despite these qualities, VAWTs are still an underutilized technology for RE production in cities. Here we argue that a major factor limiting the spread of VAWT for energy micro-generation in urban environments is the lack of quantitative information on production potential available to consumers, especially at the city-scale and in complex urban landscapes. This is not surprising though, as long-term wind measurements, commonly used in preliminary wind farm siting studies, are considerably more expensive and less effective in highly heterogeneous and aerodynamically resistant urban landscapes. The scientific literature on wind energy in urban environments mostly consists of computational fluid dynamics (CFD) simulations investigating design and efficiency of individual devices at the building and rotor scale [15,16]. Globally, only a handful of studies addressed this scientific question from a meteorological perspective at the neighborhood or city scale [17,18], and the dire need for wind resource mapping of urban areas has been highlighted by recent review works [19,20].

To address this knowledge gap, in this study we use neighborhood-scale city-wide Weather Research and Forecast (WRF) model simulations in combination with power output curves describing the performance of a set of commercially available small scale VAWTs. The objective is to produce a first-order quantitative and comparative assessment of wind energy micro-generation potential across the cities of Lausanne and Geneva, Switzerland (Figure 1).

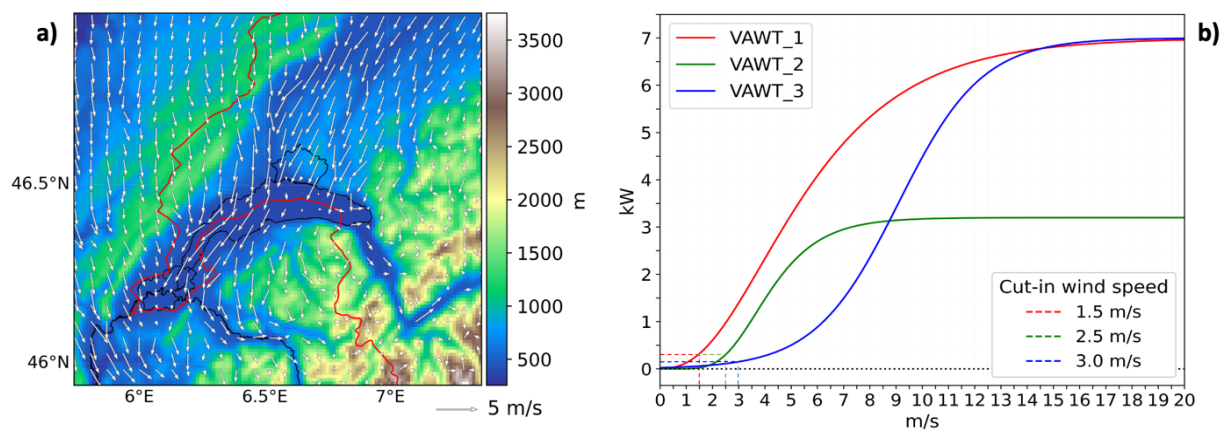

**Figure 1** Conceptual illustration of the tools used to perform the quantitative assessment of wind energy micro-generation potential. **a)** Terrain elevation (color shading) and wind flow field (white arrows) over the Léman lake (black outline) region. The municipalities of Lausanne and Geneva are represented as black outlines, while the red line illustrates the Swiss national boundary. **b)** Power curves of the three case-study VAWTs used in this study. Color coded dashed lines illustrate cut-in wind speeds and corresponding power outputs for each turbine.

We focus on these Swiss cities for two main reasons. First, because of their geographic setting as the complex topography and lake-land circulations generate a variety of wind patterns, with unclear consequences on the potential for wind energy generation. Second, because Switzerland, by voting the new Climate and Innovation Act that sets a net-zero emission target for the year 2050 [21], will be an important testbed of the green energy transition in the next 25 years. Switzerland also decided to gradually decommission the country's nuclear power plants [22], hence, such an ambitious goal will have to be achieved with the use of carbon sequestration and storage technologies and by fully transitioning the country's energy sector from fossil fuels to RE [23]. However, the space effectively available for WT deployment is limited by technical constraints (e.g., complex and rugged terrain, lakes and glaciers, sensitive infrastructures like airports and military areas), land use conflicts and social acceptance [24, 25, 26]. Thus, there is the need to assess and evaluate the potential contribution from wind energy micro-generation in urban environments.

This research study is structured as follows. Section 2 provides a detailed description of the numerical simulations performed to obtain neighborhood-scale gridded wind estimates and the calculation of potential wind power outputs. Section 3 presents model evaluation, wind speed simulation and power output calculation results at the diurnal, seasonal and annual timescales. Section 4 summarizes the findings, discusses limitations and trade-offs, and proposes suggestions and perspectives for future research.

## 2. Materials and Methods

*2.1 Wind speed numerical modeling*

We generate mean wind speed data for energy resource assessment using the Weather Research and Forecast (WRF) model version 4.5.1 [27]. WRF is an open source, highly customizable, state-of-the-art mesoscale numerical weather prediction model, created and maintained by the National Center of Atmospheric Research (NCAR), in collaboration with other United States scientific institutions. It features contributions and components developed by its worldwide community of users to address specific scientific questions, including wind energy [28] and urban climate studies [29]. WRF operates a dynamical down-scaling of coarse spatial resolution (1 geographical degree to 25 km) climatological data to resolve regional scale (10 to 0.5 km) meteorological processes, using fine resolution geographical data (e.g., topography, land cover, impervious fraction). In this study, the numerical downscaling process is performed through four one-way nested domains with increasing spatial resolution (d01: 3 km, d02: 1 km, d03 and d04: 333 m; Figure 2a) to resolve urban boundary layer (UBL) dynamics at the scale of few building blocks while ensuring consistency of the Navier-Stokes numerical solutions. Gridded climatological input, at a spatial resolution of 25 km and a temporal resolution of 6 hours, are derived from the NCEP GDAS/FNL reanalysis dataset [30]. Urban landscapes are reproduced through 10 land cover categories corresponding to the local climate zones framework (LCZ, Figure 2b and 2c) of Stewart and Oke [31]. Each LCZ corresponds to a different combination of building heights, paved surface fraction, thermal properties, and street parameters. Global LCZ maps based on Landsat images classified with a random forest algorithm, have been first published in 2018 with the creation of the World Urban Database and Access Portal Tools (WUDAPT) initiative [32]. Starting with version 4.3, the LCZ scheme is fully integrated into

WRF through the implementation of a specific table of parameters characterizing each LCZ [33] and a hybrid 100 m resolution global land cover dataset combining global maps of LCZ and Copernicus Global Land Service Lans Cover data resampled to the classification scheme of the NASA Moderate Resolution Imaging Spectroradiometer (MODIS) satellite mission [34].

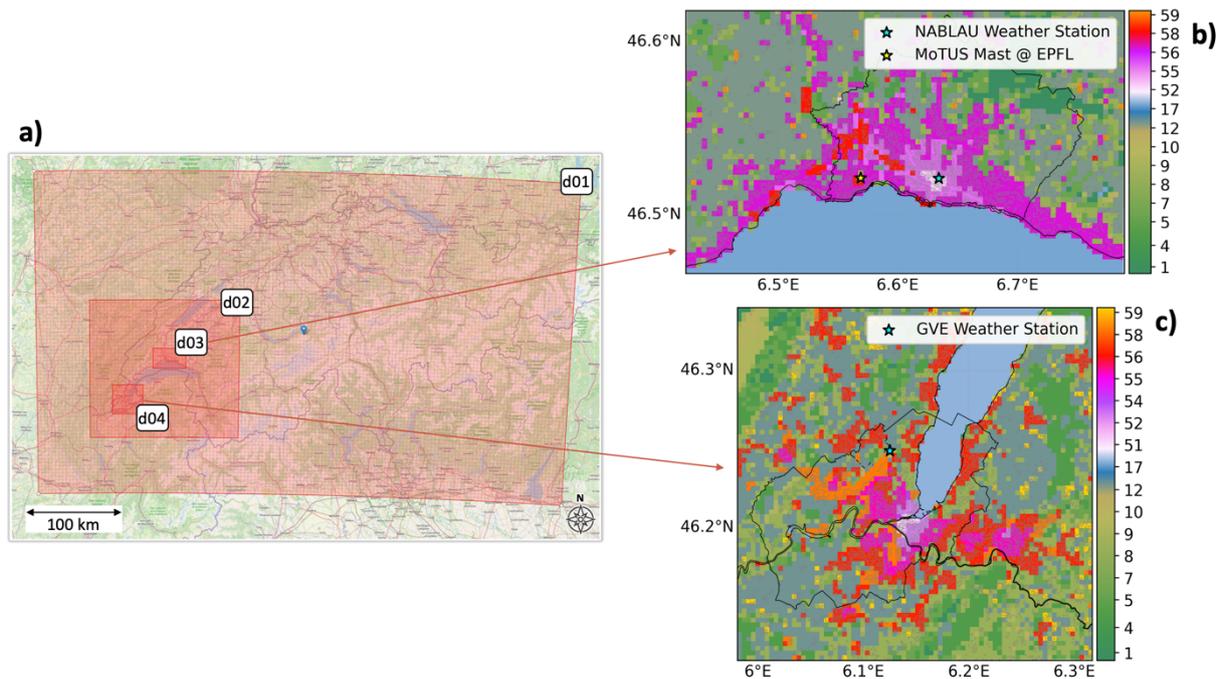

**Figure 2 a)** Simulation domains used in this study. **b)** Map of land cover representation in the d03 simulation domain over the city of Lausanne, as a combination of LCZ (51-59) in urban areas and MODIS land cover categories (1-17) across natural landscapes. **c)** As in **b)** but for d04 domain over the city of Geneva. The cyan stars in **b)** and **c)** represent the location of the NABLAU (Lausanne) and GVE (Geneva) weather stations used for model evaluation. The yellow star in **b)** represents the location of the MoTUS instrumented mast on EPFL campus, Lausanne. The LCZ represented in this study simulations are: 51 – Compact highrise (Geneva only), 52 – Compact midrise; 54 – Open highrise (Geneva only); 55 – Open midrise; 56 – Open lowrise; 58 – Large lowrise; 59 – Sparsely built. MODIS land cover categories: 1 – Evergreen Needleleaf Forest; 4 – Deciduous Broadleaf Forest; 5 – Mixed Forest; 7 – Open Shrublands; 8 – Woody Savannas; 9 – Savannas; 10 – Grasslands; 12 – Croplands; 17 – Water.

The vertical space is discretized in 85 model layers, with the lowest one having a depth of 50 m and a center of mass (where 3D meteorological variables, including wind speed, are calculated) at 25 m above ground level (AGL). Since the height of most Swiss buildings is estimated to be between 9 and 15 m [35], the center of mass of the lowest model layer is situated within the lower inertial sublayer, i.e., 1.5-3 times the height of urban elements [10]. This configuration allows for a calculation of mean wind speed that accounts for the impact of individual roughness elements on air flow, without misrepresenting sub-grid turbulent effects that cannot be captured by a mesoscale model like WRF. The dynamical interaction between urban land surfaces and the atmosphere is calculated by the coupled Building Effect Parameterization and Building Energy Model [36] which explicitly accounts for the impact of buildings on wind flow. The interaction between non-built landscapes and the overlying atmosphere is computed by the community Noah land surface model with multiparameterization options (Noah-MP) which explicitly accounts for vegetation phenology and canopy energy balance, as well as snow cover and soil hydrology [37]. Both land surface models (LSM) rely on the Janjic Eta surface parameterization scheme and are coupled to the troposphere through the Mellor-Yamada-Janjic (MYJ) planetary boundary layer scheme [38]. To attain a meaningful assessment of wind energy resources that accounts for seasonal variability on wind speed, we simulate the regional meteorology for the entire year of 2022, through twelve separate monthly realizations. Output from the innermost domains (d03 and d04) is saved at hourly frequency in order to resolve in detail the diurnal meteorological cycle, while also complying with computational capacity and storage space limitations. Output from the intermediate domain (d02) and outermost domain (d01) is saved every 4 and 24 hours, respectively.

*2.2 Case-study wind turbines and power output calculation*

The calculation of power output potential is performed by means of three hypothetical power curves, which closely reproduce those of VAWTs currently available on the retail market (Figure 1b). The case-study VAWTs used as reference in this study are characterized by different sizes, weights, blade designs, installation requirements, and purchase prices representing the range of possible applications. VAWT_1 is a *large* sized high-budget (several hundred kilograms, ~70k CHF) turbine featuring a 5 m tall, 3.1 m wide, helix shaped 3-blade rotor installed on a mast whose height varies from 3.5 to 15 m depending on whether the turbine is installed on a roof or on the ground. It is then assumed to be an ideal option for public institutions (e.g., town halls, sport centers and universities) or private companies that have access to ample outdoor space and funding. According to manufacturer specifications, VAWT_1 features a cut-in wind speed of 1.5 m/s, a safety cut-off wind speed of 20 m/s, and a maximum power output of 7 kW at 16 m/s. VAWT_2 is a roof-mounted mid-size mid-budget (250 kg, ~8k CHF) turbine featuring a 4 m tall, 3 m wide, H-Darrieus 3-blade rotor, suitable for large town homes and residential buildings. Manufacturer specifications include a 2.5 m/s cut-in wind speed and a maximum electrical output of 3.2 kW at 10 m/s. VAWT_3 is a small-size low-budget (25 kg, < 2k CHF) turbine featuring a modular generator that can be combined with different rotors. In our study we consider the "*low wind*" configuration, consisting of a 1 m by 1 m Savonius 3-blade rotor with a 3 m/s cut-in wind speed and maximum generation potential of 7 kW at 16 m/s. We identify this device as an adequate option for single family houses and small buildings in general.

We clarify that, in the context of our analysis, all differences among the case-study VAWTs described above are reflected solely by their representative power curves and cut-in wind speeds (Figure 1b). Power generation calculations are performed using wind speed as simulated by WRF at the center of mass of the lowest model level (i.e., at 25 m AGL), and do not reflect differences in potential installation configurations (e.g., on masts, rooftops or balconies). The power curves used in this study are based on figures publicly disseminated by the VAWT manufacturers and are mathematically represented by a re-parameterized form of the generalized logistic function [39]:

$$E = \frac{A}{1+ Qe^{-k(w-W_i)^{1/Q}}} \quad (1)$$

where $E$ is the electrical output in kW, $A$ is the maximum nominal output in kW, $Q$ is a parameter, $k$ is the maximum relative growth rate (or the curve slope at the inflection point), $w$ is wind speed in m/s and $W_i$ is the wind speed at the inflection point. Eq. (1) is applied, at each hourly output value, to all model cells classified as urban (i.e., LCZ = 51-59) and the calculation is performed only for values of wind speed greater than the cut-in wind speed values of each VAWT.

*2.3 Model evaluation data*

We evaluate our simulation results against ground observations from two weather stations of the MeteoSwiss measuring networks [40], one in Lausanne (NABLAU, 46.52 N 6.64 E) and one in Geneva (GVE, 46.25 N 6.13 E). We compare hourly values of 2 m air temperature (T2) and 10 m wind speed (WS) from the aforementioned weather stations with those simulated by WRF at the corresponding model cells (cyan stars in Figures 2b and 2c).

Due to the hourly resolution of simulated wind speed data used in this study, the results presented in the next section are calculated on the assumption that the output values of instantaneous wind speed are representative for the entire hour. Electrical output values are then expressed in kWh units. Real life wind pressure is far from being steady and is instead always characterized by turbulence, gusts, and quiet periods. To test the robustness of our assumption we compare wind speed simulation results with hourly averages of 20 Hz sonic anemometers wind measurement data collected during March of 2022 at the Urban Microclimate Measurement Mast (MoTUS, [41]) installed on the EPFL campus, in Lausanne, Switzerland (yellow star in Figure 2b).

## 3. Results and Discussion

*3.1 Evaluation results*

Figure 3 shows observed (black lines) and simulated (red lines) T2 and WS hourly values for the first 4 weeks of March 2022. Our configuration of the WRF model (see Section 2.1) proves effective in reproducing the timing and amplitude of the diurnal cycles of air temperature and wind speed at both observation locations during the evaluation period. Evaluation results are overall better for Lausanne (RMSE: T2 = 1.4 K, WS = 0.68 m/s) than for Geneva (RMSE: T2 = 2.98 K, WS = 1.28 m/s), especially for what concerns T2. Such a difference can be attributed to the positioning of the two weather stations. The Lausanne NABLAU weather station is located at the southeast edge of a concrete terrace overlooking a major road in a central part of the town, an environment properly represented by the LCZ 52 land cover category (Compact midrise, 0.99 % urban fraction). Whereas, the Geneva GVE weather station, although being located within the Cointrin airport complex, is positioned on a large patch of grass that has a much lower heat storage capacity compared to other typical airport surface materials (i.e., concrete and tarmac). In our simulations, the location of the Geneve GVE weather station corresponds to a model cell represented by the LCZ 58 (Large lowrise, 0.85% urban fraction), a reasonable land cover category for an airport. However, the mismatch between real world and simulated environments results in the model nighttime warm bias shown in Figure 3c. During the rest of the diurnal cycle, both the timing and magnitude of 2 m air temperature are correctly reproduced in our simulation.

For both cities, despite accurately reproducing the diurnal cycle of wind speed most of the time, WRF appears to slightly underestimate wind speed magnitude when compared to observations from the NABLAU and Cointrin weather stations (Figure 3b and 3d). Instead, when comparing WRF mean wind speed values and sonic anemometer measurements from the MoTUS instrumented mast on EPFL campus in Lausanne, such an underestimation is only present during the evening (from 7:00 PM to 12:00 AM local time), whereas during the rest of the diurnal cycle WRF slightly overestimates mean wind speed. Figure 4 shows the 18-day averages (12-30 March 2022) of 20 Hz sonic anemometer data (red line), hourly averaged sonic data (black dots) and hourly mean wind simulated by WRF at the corresponding model cell (blue triangles). Despite a slight model overestimation of mean wind speed during portions of the diurnal cycle, a root mean square error of 0.34 m/s demonstrates that wind speed data as simulated by our configuration of the WRF model are representative of the actual wind pressure affecting the urban areas in the region during the year 2022, adding confidence to the robustness of our methodology and to the representativeness of our results.

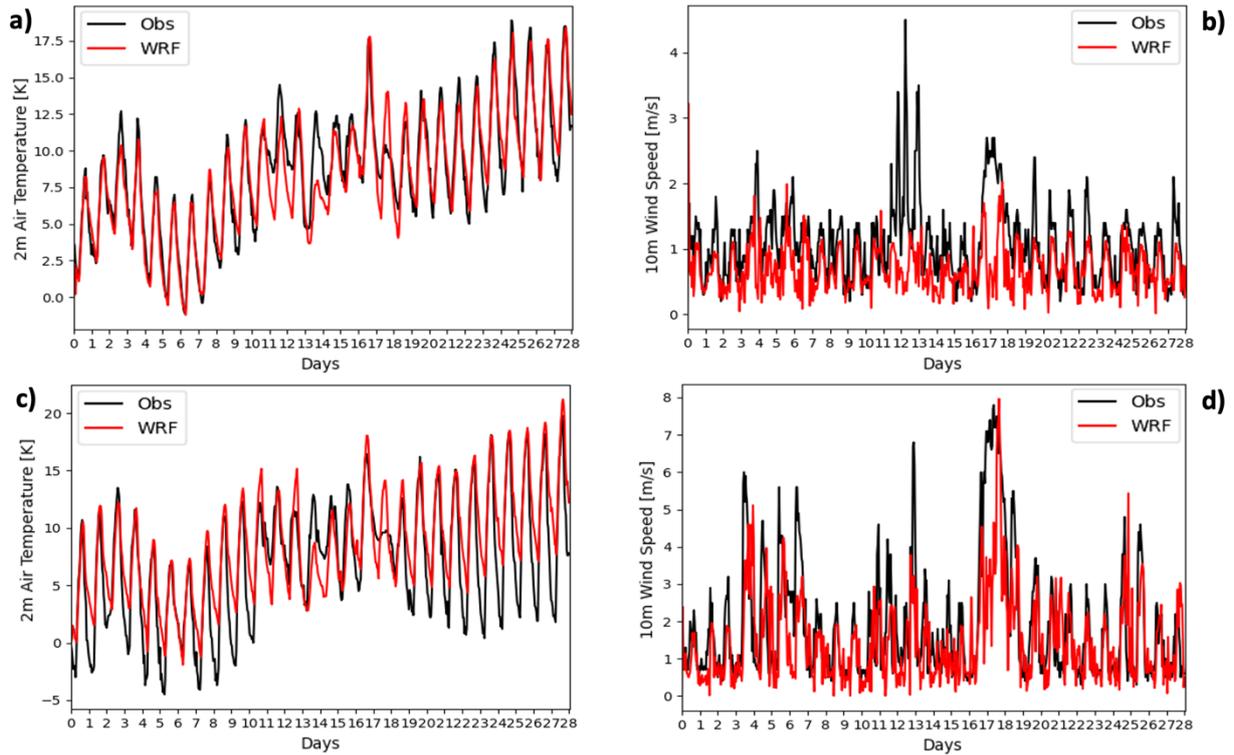

**Figure 3** Time series of **a)** 2 m air temperature and **b)** 10 m wind speed observed at the NABLAU (downtown Lausanne) weather station (black line) and simulated (red line) at the corresponding model cell, during the period 1-30 March 2022. **c)** and **d)** as in **a)** and **b)** but for GVE (Geneva airport) weather station and corresponding model cell.

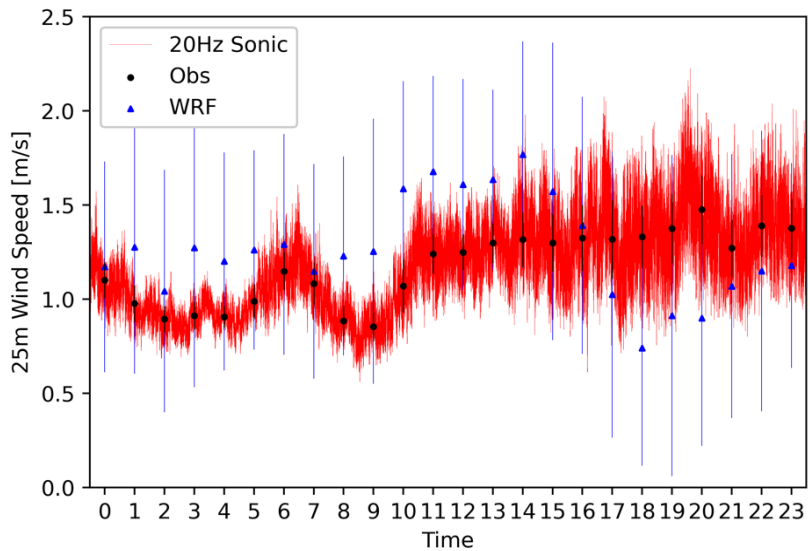

**Figure 4** Time series of 12-30 March 2022 averages of: 25 m AGL 20 Hz sonic anemometer wind speed observations (red line), hourly averaged wind speed sonic data (black dots) and simulated (red line) 25 m wind speed at the corresponding model cell. Vertical black and blue lines represent $1\sigma$ standard deviation of the temporal average of observed and simulated data, respectively.

*3.2 Wind speed in Lausanne and Geneva*

Figure 5 shows monthly averages of simulated mean wind speed across the entire domains d03 (blue dashed line in Figure 5a) and d04 (red dashed line in Figure 5b), as well as throughout the urban model cells of Lausanne (green line in Figure 5a) and Geneva (purple line in Figure 5b), during the year 2022. On average, monthly mean wind speed is 32% higher in the d03 domain compared to the d04 domain, while across urban model cells wind speed is 24% higher in Lausanne than in Geneva (Supplementary Table I). The only exception consists of the months of May and June 2022, when mean wind speed averaged across the urban cells of both cities is comparable at 2.06 m/s and 2.20 m/s, respectively.

Mean wind speed patterns also differ between Lausanne and Geneva throughout the year. In the Lausanne region (domain d03), the annual cycle of average mean wind speed is characterized by two high wind ridges during summertime and the wintertime, and two low wind troughs in March, and during the months of October and November (Figure 5a). Conversely, in the Geneva region (domain d04) there is only one high wind period corresponding to the warm season going from April to September (AMJJAS), whereas average mean wind speed remains consistently lower than 2 m/s during the rest of the year 2022 (Figure 5b). Such differences in mean wind speed are mostly attributable to the geographic location of the two cities. Lausanne resides at the southern end of the Swiss plateau, a foreland region between the Alps to the east and the Jura mountains to the west, were synoptic winds coming from central Europe often get channeled into and interact with the local topography. A typical example of this synoptic impact on local wind phenomenology is the Bise, a postfrontal regional wind typical of the springtime that often reappears during the winter [42] and is responsible for the significant wind speed differences between Lausanne and Geneva during the cold season (DJF, Figure 5).

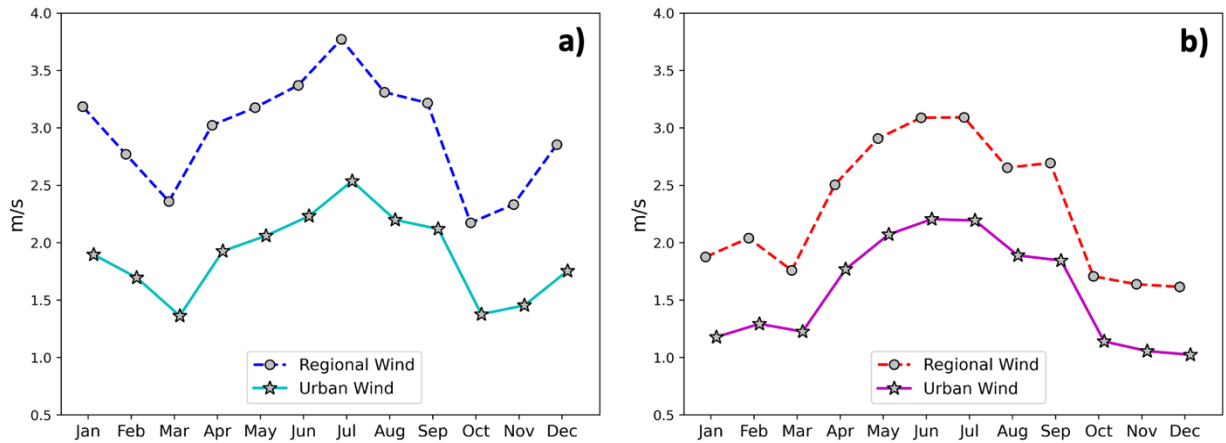

**Figure 5 a)** Time series of simulated monthly averages of mean wind speed across the entire d03 domain (blue dashed line) and the Lausanne metropolitan area (green solid line, urban model cells only). **b)** Time series of simulated monthly averages of mean wind speed across the entire d04 domain (red dashed line) and the Geneva metropolitan area (purple solid line, urban model cells only).

The geographical location of the two cities also affects their diurnal cycles of mean wind speed. During the winter (DJF) and the fall (SON) seasons the diurnal evolution of mean wind speed of both cities is characterized by higher winds at nighttime and lower winds during the daytime as a consequence of the katabatic flows coming from the Jura mountains to the northwest, the Swiss plateau to the north (Lausanne only) and the western Alps (Geneva only). During the spring (MAM) and summer (JJA) seasons, the differential surface heating between the Léman lake and the landmasses surrounding it determines the formation of pressure gradients that drive a significant local thermal circulation, in a lake breeze fashion (Figure 6b). This thermal circulation interacts with the dominant spring and summertime northerly synoptic flow, opposing it in Lausanne and enhancing it in Geneva during the daytime. At nighttime the opposite happens, as the differential heating, and the associated pressure gradient and lake breeze flow, reverse. As a result, summertime wind speed is greater at nighttime in Lausanne and during the daytime in Geneva (Figure 6a).

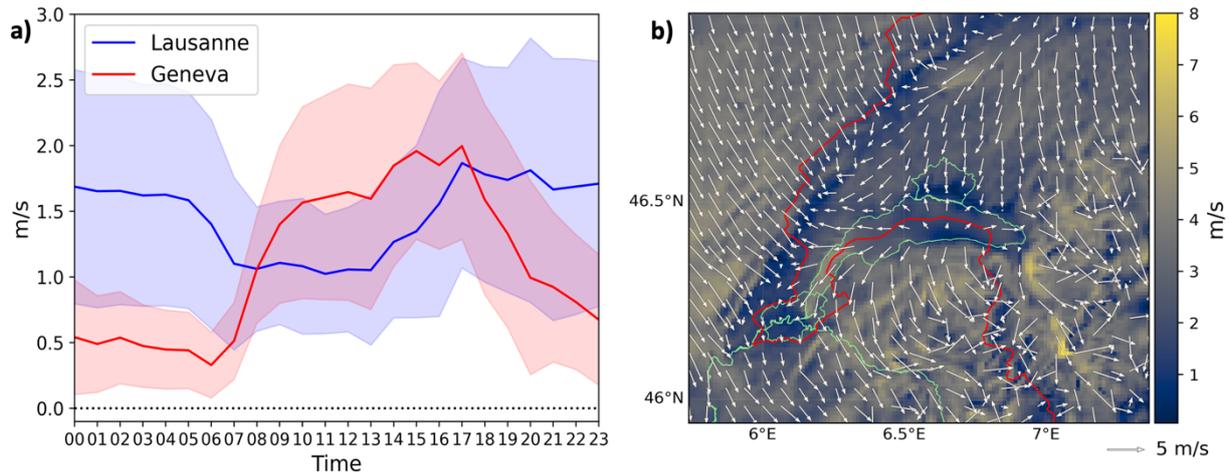

**Figure 6 a)** Time series of JJA 2022 averages of simulated hourly mean wind speed across the cities of Lausanne (blue line) and Geneva (red line). Shaded areas represent $1\sigma$ standard deviation of the temporal average. **b)** Map showing the July 2022 average of 1:00 PM local time wind vectors (white arrows) across the d02 domain. Colour shading represents mean wind speed. Solid green line represents the administrative borders of the Lausanne and Geneva metropolitan areas as well as the Léman lake shores, while the solid red line represents the Swiss state border.

*3.3 Seasonal and diurnal trends of micro-generation potentials*

The differences in mean wind speed magnitude and seasonal trends between Lausanne and Geneva described in Section 3.2 affect in turn the potential for wind energy micro-generation within the two cities throughout the year 2022 (Figure 7).

Due to the strong spring and summertime synoptic flows, both cities present the greatest micro-generation potential during the warmer half of the year 2022 (AMJJAS), in net opposition to the national non-urban wind energy trends where wintertime generation dominates [43]. During the colder half of the year (from October to March, ONDJFM), due to the difference in mean wind speed magnitude between the two cities (Section 3.2), only Lausanne presents a significant micro-generation potential (Figure 7a), as in Geneva wind speed rarely passes the VAWTs cut-in wind speed threshold.

Because of the generally higher wind speed, each of the three case-study VAWTs has the potential to individually produce on average 27.1% more electricity in Lausanne than in Geneva, across the year 2022 (Supplementary Table II, legend boxes in Figure 7). The only time when the average individual turbine generation potential is greater in Geneva than in Lausanne is during the months of May and June. This happens as the diurnal cycle of mean wind speed is considerably different between the two cities (Figure 8), despite a comparable average mean wind speed during both months. In Lausanne, mean wind speed oscillates between ~0.7 and ~1.5 m/s throughout the May and June average diurnal cycles, displaying a bi-modal two ridges/two troughs pattern (black line in Figure 8a and 8c) that closely resembles the seasonal fluctuation described in Section 3.2. Conversely, in Geneva the May and June monthly averages of hourly mean wind speed clearly display an alternation between a daytime high wind period and a nighttime low wind phase (black line in Figure 8b and 8d), here too in close resemblance to the seasonal wind speed variation in the region.

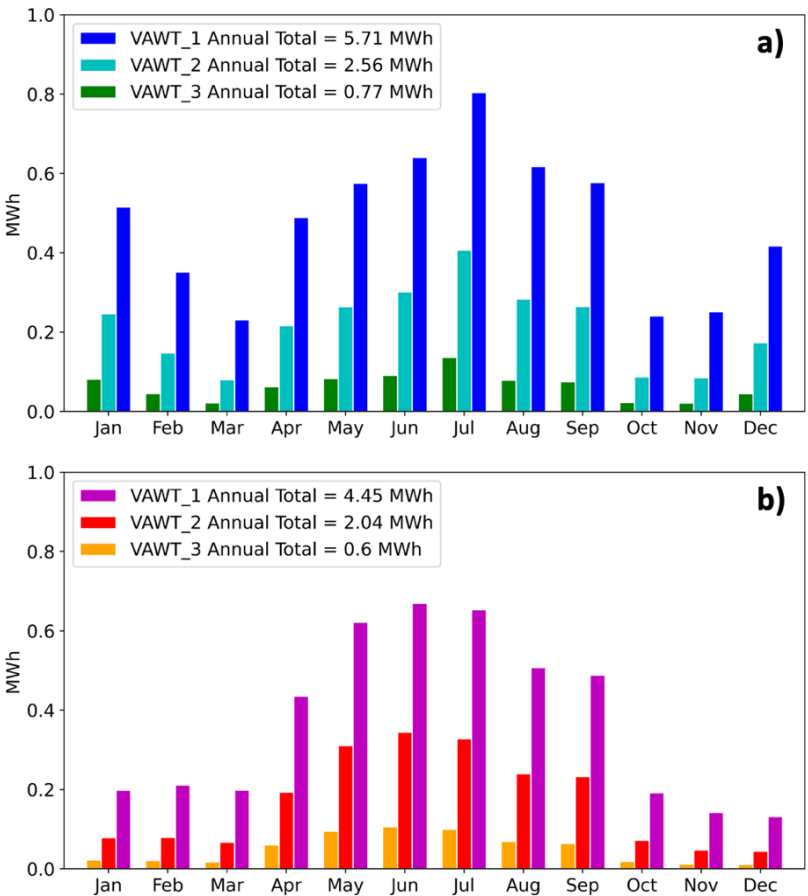

**Figure 7 a)** Time series of individual turbine monthly total wind energy micro-generation potential calculated as the average output across all urban model cells in domain d03 (Lausanne metro area) for the three case-study VAWTs. **b)** Time series of individual turbine monthly total wind energy micro-generation potential calculated as the average output across all urban model cells in domain d04 (Geneva metro area) for the three case-study VAWTs.

Such a difference in wind speed diurnal cycle between the two cities becomes extremely relevant when considering specific VAWTs cut-in wind speeds. In Lausanne, wind speed rarely crosses such thresholds during the months of May and June 2022, whereas Geneva often experiences wind speeds above 2.0 m/s during its daytime high wind period, with city average wind speed peaking at 3.5 m/s at 3:00 PM local time in June (grey shaded areas in Figure 8d). Figures 7 and 8 also show the great differences in micro-generation potential that exist among the three case-study VAWTs. Such differences are a function of each VAWT cut-in wind speed and power curve inflection points. As mean wind speed is often lower than 4.0 m/s within both urban areas, the first portion of the power curve is crucial in determining the turbine's potential power output. VAWT_1 presents monthly total generation potentials which are on average more than double in magnitude compared to VAWT_2, despite the similarity between their power curves. VAWT_3, with its 3 m/s cut-in wind speed and late inflection point, is the least performant device with monthly total power output potentials that are one order of magnitude smaller than the other two VAWTs.

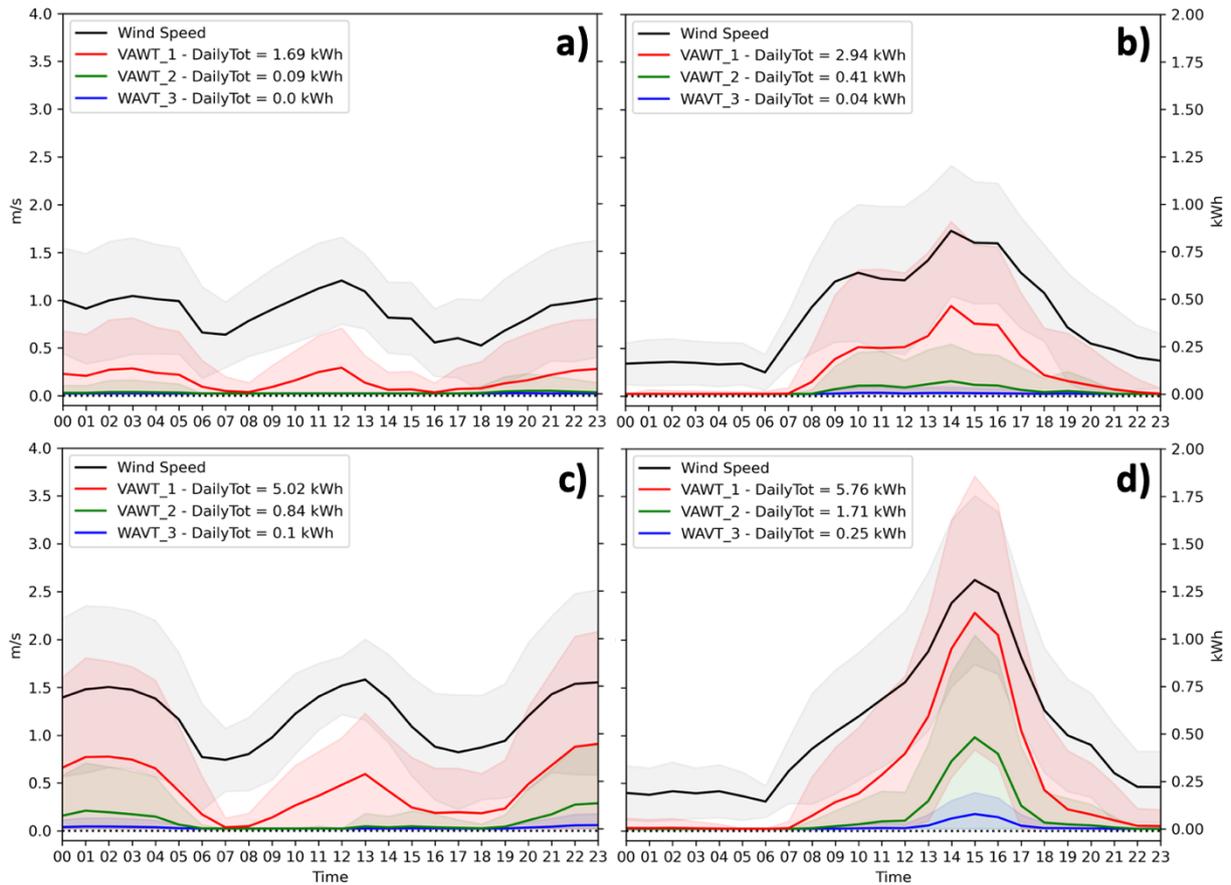

**Figure 8 a)** Time series of May 2022 monthly averages of simulated hourly wind speed (black lines) and single turbine hourly total wind energy micro-generation potential calculated as the average output across all urban model cells in domain d03 (Lausanne metro area) for the three case-study VAWTs. **b)** Time series of May 2022 monthly averages of single turbine monthly total wind energy micro-generation potential calculated as the average output across all urban model cells in domain d04 (Geneva metro area) for the three case-study VAWTs. **c)** and **d)** as in **a)** and **b)** but for the month of June 2022. Shaded areas represent $1\sigma$ standard deviation of the temporal average.

In order to provide a more practical sense of the different magnitudes in potential electrical output between the three case-study VAWTs, we compare the average individual turbine total annual potentials in both Lausanne and Geneva shown in Figure 7 with the surface averaged total annual electrical output produced in 2022 by photovoltaic panels (PVs) managed or financed by the city of Lausanne (161,7 kWh/m²y, [44]). On average, an individual VAWT in 2022 would have produced an amount of electricity equivalent to that generated by 16.5 m² of

PVs, with a great degree of variability between Lausanne and Geneva as well as between the three case-study turbines. The detailed results of the comparison are listed in Table I below.

**Table I**

*Summary of 2022 annual production equivalence between VAWTs and PVs[a].*

|        | Lausanne |      | Geneva |      |
| ------ | -------- | ---- | ------ | ---- |
|        | kWh/y    | m²   | kWh/y  | m²   |
| **VAWT_1** | 5710     | 35.3 | 4450   | 27.5 |
| **VAWT_2** | 2560     | 15.8 | 2040   | 12.6 |
| **VAWT_3** | 690      | 4.3  | 540    | 3.3  |

[a] PVs operated and financed by the city of Lausanne [42].

*3.4 Spatial distribution of potentials*

In Figures 5a and 8, we only show uncertainty of results as $1\sigma$ standard deviation of the temporal averages, which in turn is calculated from city averaged values (i.e., averaged across all urban model cells). Since we do not show the uncertainty related to spatial variability there for readability of the figures, in this subsection we show 333 m resolution spatial distribution maps of micro-generation potentials in both cities for the VAWT_1 and VAWT_2 devices (as VAWT_3 potential output is often negligible).

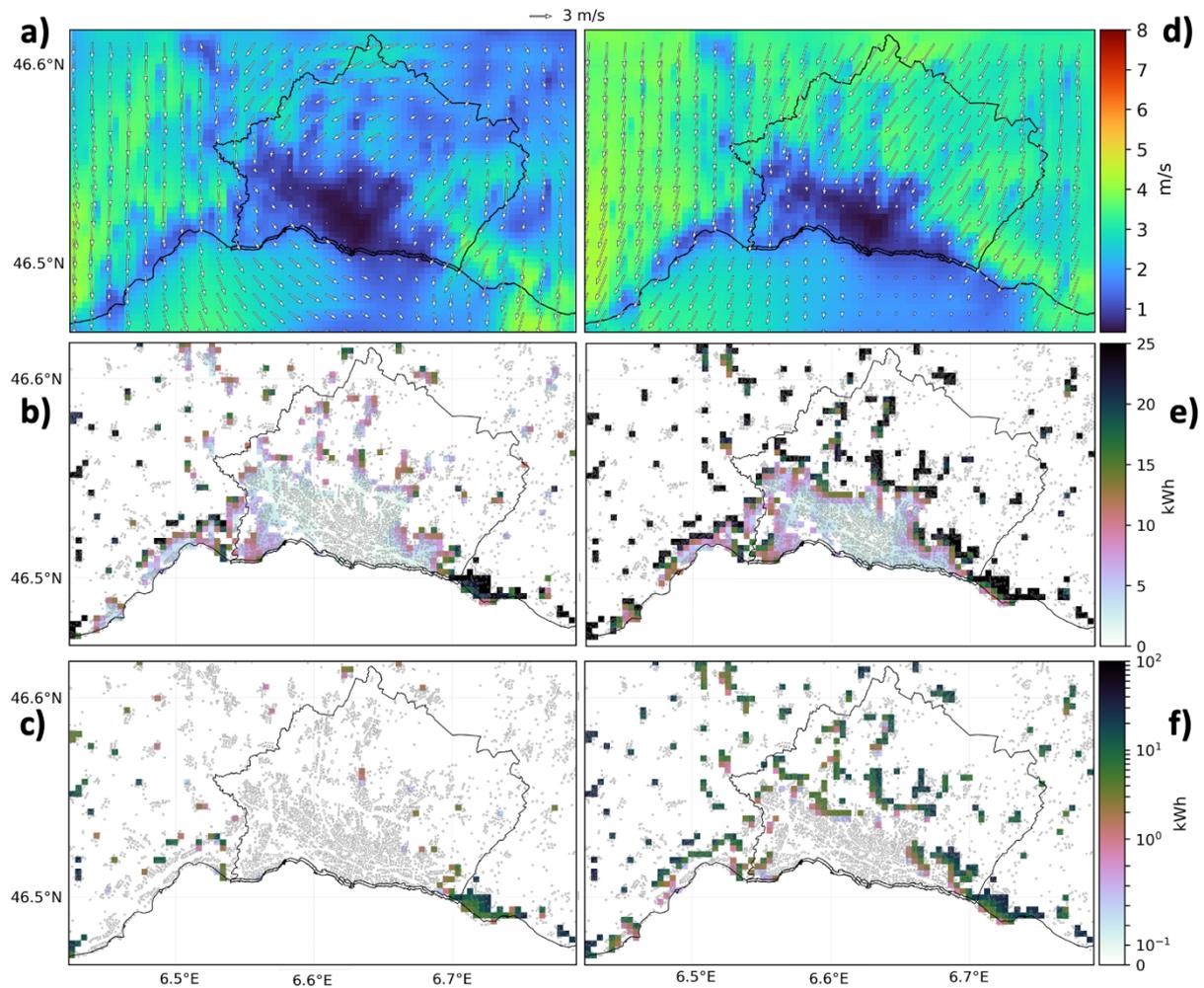

**Figure 9 a)** Map of January 2022 monthly average of mean wind speed (colour shading) and wind vectors (white arrows) across domain d03. **b)** Map of January 2022 monthly average of individual turbine daily total micro-generation potential for the VAWT_1 device. **c)** Map of January 2022 monthly average of individual turbine daily total micro-generation potential for the VAWT_2 device. **d)**, **e)** and **f)** as in **a)**, **b)** and **c)** but for the month of July 2022. The administrative boundaries of the Lausanne metropolitan area, the Léman lake coastline (black lines) and the outline of major city buildings (grey lines) are shown for reference.

In both cities, the areas with the greatest micro-generation potential are located along the city outskirts, whereas in the city center there is non-negligible potential only during the summer months for the VAWT_1 device. In Lausanne, the areas with the highest wind generation potential are the northern fringes of the city, as the wind in Lausanne is consistently northerly (Figure 9). Instead, Geneva often experiences westerly and southerly overland winds that interact with the northerly flows over the lake in modulating the city micro-generation potentials throughout the year. Figure 10 shows an example of this modulation. During the month of June 2022, the average over land wind is westerly across domain d04 and weakly northerly over the lake (Figure 10a). As a consequence, the areas of highest potential generation are located along the southwestern and southeastern fringes of the metro area, as well as along the eastern shore of the Léman lake (Figure 10b and 10c). During the month of July 2022, the wind is instead northerly everywhere, with greater speed over the lake and on the eastern portion of the domain; thus, in June 2022 the areas of greater micro-generation potential are located on the northern and eastern margins of the metro area and along both shores of the lake. In both cities, the only areas that steadily experiences high wind and generation potential throughout the year are located along limbs extending to the eastern end of the metro areas.

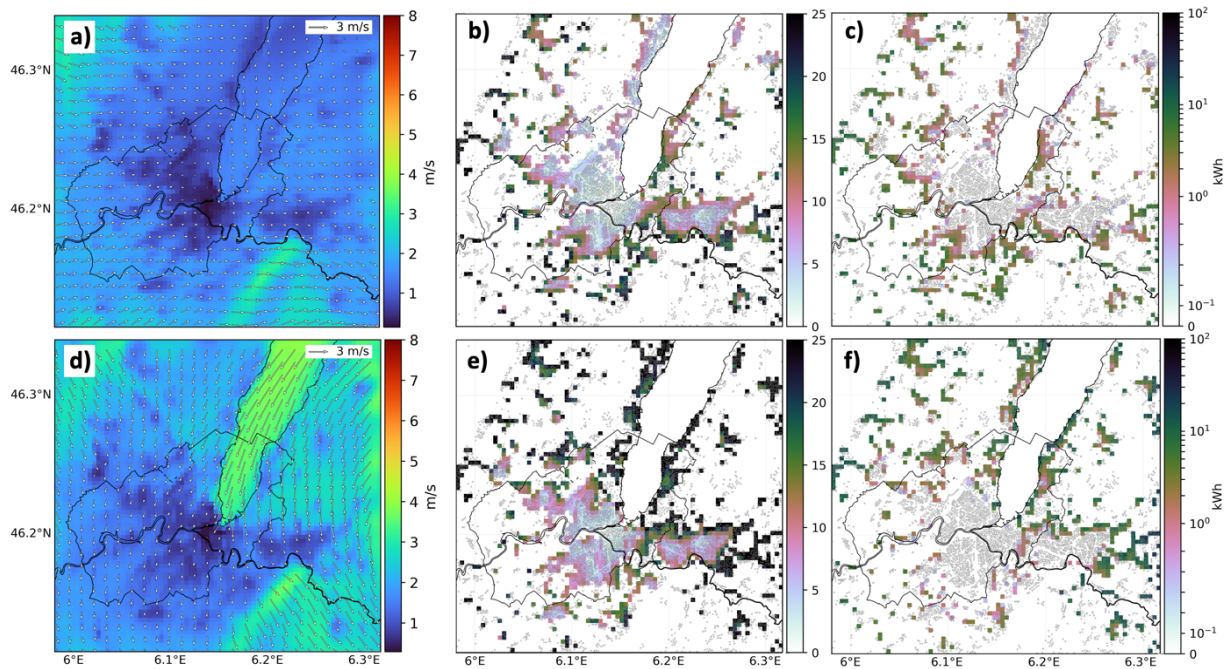

**Figure 10 a)** Map of June 2022 monthly average of mean wind speed (colour shading) and wind vectors (white arrows) across domain d03. **b)** Map of June 2022 monthly average of individual turbine daily total micro-generation potential for the VAWT_1 device. **c)** Map of June 2022 monthly average of individual turbine daily total micro-generation potential for the VAWT_2 device. **d), e)** and **f)** as in **a), b)** and **c)** but for the month of July

2022. The administrative boundaries of the Geneva metropolitan area, the Léman lake coastline (black lines) and the outline of major city buildings (grey lines) are shown for reference.

## 4. Conclusions

In the light of the 2023 Climate and Innovation Act, that prescribes a 40-fold increase in wind energy production across the national territory by 2050, Switzerland is set to face great implementation challenges due to technical complexities and public acceptance. Drawing on the need to explore all available options, and in the face of the current lack of scientific and institutional assessments of wind energy resources in urban environments, in this study we investigated the potential for distributed wind energy micro-generation in two Swiss cities. We used a set of high-resolution Weather Research and Forecast model simulations to generate hourly 25 m AGL wind speed data across the Léman lake region for the entire year of 2022. We calculated wind energy micro-generation potentials by inputting wind speed data in parametric equations that closely reproduce the power curves of VAWTs currently available on the retail market. We focused our analysis on the cities of Lausanne and Geneva to explore the role of different geographical settings, local meteorology, and urban extent on wind energy production.

Our analysis shows that diurnal and seasonal trends of mean wind speed and associated wind energy micro-generation are significantly different between the two cities, due to the dynamical interaction between synoptic flows and the local thermal circulation (i.e., lake breeze). Both cities experience absolute maximum mean wind speed and associated wind energy potential from April to September, in opposition to average non-urban national trends. In addition, in the Lausanne region a relative maximum of wind speed and energy generation is present from December to February. This is a particularly desirable characteristic, in terms of a comprehensive RE strategy, as both hydropower and photovoltaic Swiss sources experience significant production deficits during the winter, and the country heavily relies on imports to match its energy demand [45]. During the spring and summer seasons, the lake breeze circulation opposes the synoptic flow in Lausanne and enhances it in Geneva during the daytime, while the opposite happens at nighttime. As a result, during the warm season, wind speed and energy production are greater at daytime in Geneva and at nighttime in Lausanne, once again favoring the integration with other RE sources, especially photovoltaics, within the latter. In Geneva, the impact of VAWTs generation could be maximized through the implementation of tailored time-shifting demand response strategies, in a similar fashion to what suggested by Kontani and Tanaka [46].

Overall, we estimate that on average, during the year 2022, an individual VAWT would have generated an amount of electricity equivalent to 16.5 $m^2$ of PVs, resulting in a potential expansion of available surface for RE microgeneration. In both cities, the areas with greater potential for wind energy harvesting are located along the external fringes of the conurbations. These results prompt for careful planning strategies to be implemented in conjunction with any urban wind energy effort as future urban outward expansion may reduce the power generation of currently attractive areas. In Lausanne, where winds are predominantly northerly throughout the year, these areas are located on the northern side and lateral edges of the city. Geneva instead, often experiences westerly and southerly overland winds that interact with the northerly winds over the lake. This interaction determines a shift in wind energy potential between the southern and northern fringes of the city throughout the year. In both city centers, only the VAWT_1 case-study wind turbine is able to generate significant energy outputs during the summer months, by virtue of its low cut-in wind speed and early power curve inflection point. However, large portions of both city centers are included in the Swiss Federal Inventory of

Heritage Sites of national importance (ISOS, [47]) which strongly limits the extent and nature of allowed modifications to buildings, including the installation of wind turbines.

From our analysis, it clearly emerges that great technical improvement in VAWT design is needed, especially with regards to efficiency at low wind regimes. Several commercially available small size VAWTs boast potential maximum electrical outputs of 5 to 7 kW (e.g., VAWT_3 in this study), despite the fact that the wind speeds required to attain such a performance are rarely, if ever, attained within urban environments. Here we find that the most crucial features for meaningful wind energy micro-generation are low cut-in wind speeds (ideally < 1 m/s) and early inflection points, thus highlighting the need for engineering research on small size VAWTs to prioritize low wind efficiency over maximum potential output in order to allow for a significant contribution to RE generation in urban environments.

Due to the regional and city-wide scale of our study, and the limitations associated with computational power and data storage capacity, we used a mesoscale meteorological model (WRF) that does not resolve micro-scale turbulence and wake effects generated by highly complex urban morphological features in the roughness sublayer heavily affecting WT efficiency and performance [48]. For this reason, we chose to focus our analysis within the lower portion of the inertial sub-layer (i.e., above the blending height) where the impact of roughness elements is horizontally homogenized [10]. However, we acknowledge that individual turbine placement and surrounding morphological features, like roof geometry and trees, are expected to sensibly affect the downscaling of our results. The next stage of this research is to use the meteorological output from this study to force finer-scale simulations (e.g., CFD) and dynamically resolve micro-scale flow characteristic within and aloft the urban canopy.

With this study we contribute to filling a gap in the scientific literature on wind energy resources in urban areas, by providing a first order, city-wide, numerical assessment of micro-generation potentials in two Swiss cities. Future research across different cities and periods of study are needed to expand and improve the understanding of wind harvesting potential in urban environments.

## Acknowledgements

**Author Contributions**
**Aldo Brandi:** Conceptualization, Methodology, Software, Formal analysis, Investigation, Data curation, Writing – Original Draft, Visualization. **Gabriele Manoli:** Formal analysis, Resources, Writing – Review & Editing.

**Funding**
This research did not receive any specific grant from funding agencies in the public, commercial, or not-for-profit sectors.

**Data availability**
The data that support the findings of this study are available from the corresponding author upon reasonable request.

# References

[1] Seto, K. C., Dhakal, S., Bigio, A., Blanco, H., Carlo Delgado, G., Dewar, D., ... & Zwickel, T. (2014). *Human settlements, infrastructure, and spatial planning*.

[2] Seto, K. C., Churkina, G., Hsu, A., Keller, M., Newman, P. W., Qin, B., & Ramaswami, A. (2021). From low- to net-zero carbon cities: The next global agenda. *Annual review of environment and resources*, *46*, 377-415.

[3] Kii, M. (2021). Projecting future populations of urban agglomerations around the world and through the 21st century. *npj Urban Sustainability*, *1*(1), 10.

[4] Rahman, A., Farrok, O., & Haque, M. M. (2022). Environmental impact of renewable energy source based electrical power plants: Solar, wind, hydroelectric, biomass, geothermal, tidal, ocean, and osmotic. *Renewable and Sustainable Energy Reviews*, *161*, 112279.

[5] World Wind Energy Association (WWEA) Half-year Report 2023 : Additional Momentum for Windpower in 2023 (2023) https://wwindea.org/wwea-half-year-report-2023-additional-momentum-for-windpower-in-2023

[6] Grawe, D., Thompson, H. L., Salmond, J. A., Cai, X. M., & Schlünzen, K. H. (2013). Modelling the impact of urbanisation on regional climate in the Greater London Area. *International Journal of Climatology*, *33*(10), 2388-2401.

[7] McVicar, T. R., Roderick, M. L., Donohue, R. J., Li, L. T., Van Niel, T. G., Thomas, A., ... & Dinpashoh, Y. (2012). Global review and synthesis of trends in observed terrestrial near-surface wind speeds: Implications for evaporation. *Journal of Hydrology*, *416*, 182-205.

[8] Rafailidis, S. (1997). Influence of building areal density and roof shape on the wind characteristics above a town. *Boundary-layer meteorology*, *85*, 255-271.

[9] Bornstein, R. D., & Johnson, D. S. (1977). Urban-rural wind velocity differences. *Atmospheric Environment (1967)*, *11*(7), 597-604.

[10] Oke, T. R., Mills, G., Christen, A., & Voogt, J. A. (2017). *Urban climates*. Cambridge University Press.

[11] Fortuniak, K., Kłysik, K., & Wibig, J. (2006). Urban–rural contrasts of meteorological parameters in 648 Łódź. Theoretical and applied climatology, 84, 91-101.

[12] Droste, A. M., Steeneveld, G. J., & Holtslag, A. A. (2018). Introducing the urban wind island effect. *Environmental Research Letters*, *13*(9), 094007.

[13] Stewart, I. D., & Mills, G. (2021). *The urban heat Island*. Elsevier.

[14] Fernando, H. J. S. (2010). Fluid dynamics of urban atmospheres in complex terrain. *Annual review of fluid mechanics*, *42*, 365-389.

[15] Toja-Silva, F., Kono, T., Peralta, C., Lopez-Garcia, O., & Chen, J. (2018). A review of computational fluid dynamics (CFD) simulations of the wind flow around buildings for urban wind energy exploitation. *Journal of Wind Engineering and Industrial Aerodynamics*, *180*, 66-87.

[16] Tan, J. D., Chang, C. C. W., Bhuiyan, M. A. S., Nisa'Minhad, K., & Ali, K. (2022). Advancements of wind energy conversion systems for low-wind urban environments: A review. *Energy Reports*, *8*, 3406-3414.

[17] Simões, T., & Estanqueiro, A. (2016). A new methodology for urban wind resource assessment. *Renewable Energy*, *89*, 598-605

[18] Mi, L., Han, Y., Shen, L., Cai, C., & Wu, T. (2022). Multi-Scale Numerical Assessments of Urban Wind Resource Using Coupled WRF-BEP and RANS Simulation: A Case Study. *Atmosphere*, *13*(11), 1753.

[19] Anup, K. C., Whale, J., & Urmee, T. (2019). Urban wind conditions and small wind turbines in the built environment: A review. *Renewable energy*, *131*, 268-283.

[20] Tasneem, Z., Al Noman, A., Das, S. K., Saha, D. K., Islam, M. R., Ali, M. F., ... & Alam, F. (2020). An analytical review on the evaluation of wind resource and wind turbine for urban application: Prospect and challenges. *Developments in the Built Environment*, *4*, 100033.

[21] Climate and Innovation Act. https://www.admin.ch/gov/en/start/documentation/votes/20230618/climate-and-innovation-act.html Accessed 18 May 2024.

[22] Changes in the Law of Nuclear Energy. https://www.bfe.admin.ch/bfe/en/home/policy/energy-strategy-2050/initial-package-of-measures/changes-in-the-law-on-nuclear-energy.html Accessed 18 May 2024.

[23] Energy Perspective 2050. https://www.bfe.admin.ch/bfe/en/home/policy/energy-strategy-2050/documentation/energy-perspectives-2050.html Accessed 18 May 2024.

[24] Where should wind turbines be constructed in Switzerland? https://ethz.ch/en/news-and-events/eth-news/news/2023/03/where-should-wind-turbines-be-constructed-in-switzerland.html Accessed 18 May 2024.

[25] Vuichard, P., Broughel, A., Wüstenhagen, R., Tabi, A., & Knauf, J. (2022). Keep it local and bird-friendly: Exploring the social acceptance of wind energy in Switzerland, Estonia, and Ukraine. *Energy Research & Social Science*, *88*, 102508.

[26] Broughel, A., & Wüstenhagen, R. (2022). The influence of policy risk on Swiss wind power investment. *Swiss Energy Governance: Political, Economic and Legal Challenges and Opportunities in the Energy Transition*, 345-368.

[27] Skamarock, W. C., Klemp, J. B., Dudhia, J., Gill, D. O., Liu, Z., Berner, J., ... Huang, X. -yu. (2021). A Description of the Advanced Research WRF Model Version 4.3 (No. NCAR/TN-556+STR). doi:10.5065/1dfh-6p97

[28] Al-Yahyai, S., Charabi, Y., & Gastli, A. (2010). Review of the use of numerical weather prediction (NWP) models for wind energy assessment. *Renewable and Sustainable Energy Reviews*, *14*(9), 3192-3198.

[29] Zhu, D., & Ooka, R. (2023). WRF-based scenario experiment research on urban heat island: A review. *Urban Climate*, *49*, 101512.

[30] National Centers for Environmental Prediction/National Weather Service/NOAA/U.S. Department of Commerce. 2015, updated daily. *NCEP GDAS/FNL 0.25 Degree Global Tropospheric Analyses and Forecast Grids*. Research Data Archive at the National Center for Atmospheric Research, Computational and Information Systems Laboratory. https://doi.org/10.5065/D65Q4T4Z. Accessed 18 May 2024.

[31] Stewart, I. D., & Oke, T. R. (2012). Local climate zones for urban temperature studies. *Bulletin of the American Meteorological Society*, *93*(12), 1879-1900.

[32] Ching, J., Mills, G., Bechtel, B., See, L., Feddema, J., Wang, X., ... & Theeuwes, N. (2018). WUDAPT: An urban weather, climate, and environmental modeling infrastructure for the anthropocene. *Bulletin of the American Meteorological Society*, *99*(9), 1907-1924.

[33] Zonato, A., & Chen, F. (2021). Updates of WRF-urban in WRF 4.3: Local Climate Zones, Mitigation Strategies, building materials permeability and new buildings drag coefficient.

[34] Demuzere, M., He, C., Martilli, A., & Zonato, A. (2023). Technical documentation for the hybrid 100-m global land cover dataset with Local Climate Zones for WRF.

[35] Buildings and Dwellings Statistics 2022. https://www.bfs.admin.ch/bfs/en/home/statistics/construction-housing/buildings/size.gnpdetail.2023-0442.html Accessed 18 May 2024.

[36] Salamanca, F., Krpo, A., Martilli, A., & Clappier, A. (2010). A new building energy model coupled with an urban 729 canopy parameterization for urban climate simulations—part I. formulation, verification, and sensitivity 730 analysis of the model. Theoretical and applied climatology, 99, 331-344.

[37] Niu, G. Y., Yang, Z. L., Mitchell, K. E., Chen, F., Ek, M. B., Barlage, M., ... & Xia, Y. (2011). The community Noah land surface model with multiparameterization options (Noah-MP): 1. Model description and evaluation with local-scale measurements. *Journal of Geophysical Research: Atmospheres*, *116*(D12).

[38] Mellor, G. L., & Yamada, T. (1982). Development of a turbulence closure model for geophysical fluid 693 problems. Reviews of Geophysics, 20(4), 851-875.

[39] Tjørve, E., & Tjørve, K. M. (2010). A unified approach to the Richards-model family for use in growth analyses: why we need only two model forms. *Journal of theoretical biology*, *267*(3), 417-425.

[40] Measurement values at meteorological stations. https://www.meteoswiss.admin.ch/weather/measurement-values-and-satellite-images/measurement-values-at-meteorological-stations.html Accessed 18 May 2024.

[41] MoTUS Archive Explorer. https://motus.epfl.ch/data/ Accessed 18 May 2024.

[42] Wanner, H., & Furger, M. (1990). The Bise—Climatology of a regional wind north of the Alps. *Meteorology and Atmospheric Physics*, *43*, 105-115.

[43] Wind Energy. https://www.bfe.admin.ch/bfe/en/home/supply/renewable-energy/wind-energy.html Accessed 18 May 2024.

[44] RAPPORT DU CONSEIL D'ETAT AU GRAND CONSEIL https://www.vd.ch/fileadmin/user_upload/organisation/gc/fichiers_pdf/2022-2027/24_RAP_18_TexteCE.pdf Accessed 18 May 2024.

[45] Lienhard, N., Mutschler, R., Leenders, L., & Rüdisüli, M. (2023). Concurrent deficit and surplus situations in the future renewable Swiss and European electricity system. *Energy Strategy Reviews*, *46*, 101036.

[46] Kontani, R., & Tanaka, K. (2024). Integrating variable renewable energy and diverse flexibilities: Supplying carbon-free energy from a wind turbine to a data center. *Urban Climate*, *54*, 101843.

[47] Federal Inventory of Heritage Sites of national importance ISOS and protection of heritage sites https://www.bak.admin.ch/bak/en/home/baukultur/isos-und-ortsbildschutz.html Accessed 18 May 2024.

[48] Fan, X., Ge, M., Tan, W., & Li, Q. (2021). Impacts of coexisting buildings and trees on the performance of rooftop wind turbines: An idealized numerical study. *Renewable Energy*, *177*, 164-180.